\documentclass[english]{article}
\usepackage[T1]{fontenc}
\usepackage[latin9]{inputenc}
\usepackage[a4paper]{geometry}
\usepackage{cite}
\usepackage{babel}
\usepackage{color}
\usepackage{array}
\usepackage{url}
\usepackage{multirow}
\usepackage{amsmath}
\usepackage{graphicx}
\usepackage{setspace}
\makeatletter

\providecommand{\tabularnewline}{\\}

\DeclareMathOperator{\rmd}{d}
\DeclareMathOperator{\Dir}{Dir}

\makeatother

\begin{document}

\title{Modeling of the parties' vote share distributions}

\author{A. Kononovicius}

\date{Institute of Theoretical Physics and Astronomy, Vilnius University}
\maketitle
\begin{abstract}
Competition between varying ideas, people and institutions fuels the
dynamics of socio-economic systems. Numerous analyses of the empirical
data extracted from different financial markets have established a
consistent set of stylized facts describing statistical signatures
of the competition in the financial markets. Having an established
and consistent set of stylized facts helps to set clear goals for
theoretical models to achieve. Despite similar abundance of empirical
analyses in sociophysics, there is no consistent set of stylized facts
describing the opinion dynamics. In this contribution we consider
the parties' vote share distributions observed during the Lithuanian
parliamentary elections. We show that most of the time empirical vote
share distributions could be well fitted by numerous different distributions.
While discussing this peculiarity we provide arguments, including
a simple agent-based model, on why the beta distribution could be
the best choice to fit the parties' vote share distributions.
\end{abstract}

\section{Introduction}

There are numerous ways every individual can be unique. Some of the
personal degrees of freedom are wholly predetermined at birth and
are not influenced by the socio-economic or cultural context, e.g.,
biological features like skin color. While some of them could be referred
to as the economic, social or cultural variables as they are not strictly
predetermined and may change to different extent depending on individual's
behavior and personal experiences, e.g., wealth, influence, religion
or political affiliation. Most of these variables are extremely interesting,
from the perspective of social science and sociophysics, as their
dynamics reflect the ongoing competition between varying ideas, people
and institutions \cite{Axelrod1997JConfRes,Macy2002ARS,Ausloos2007EPL,Akerlof2009Princeton}.

Development of simple agent-based models of socio-economic and cultural
interactions attracted a lot of interest from physicists in the recent
couple of decades \cite{Castellano2009RevModPhys,Abergel2013NEW}.
While these models can be well used as a controlled testing ground
for qualitative theories from the social sciences \cite{Axelrod1997Comp,Macy2002ARS,Conte2012EPJ,Duggins2017JASSS}.
They are also of utmost interest to physicists as these models exhibit
different complex dynamical and statistical phenomena, reminiscent
of the complex phenomena, such as phase transitions \cite{Stanley1971OUP,Castellano2000PRL},
dissipative structures \cite{Prigogine1993,Weiss2012} or non-extensive
thermodynamics \cite{Tsallis2009Springer,Ruseckas2015PhysLettA},
common in statistical physics. This interest is well grounded in the
empirical data too, as complex statistical and dynamical patterns
are observed in the empirical data from socio-economic systems as
well \cite{Conte2012EPJ,Barabasi2015CUP,Tsallis2017Entropy}. While
there is a well established set of stylized facts for the financial
markets, e.g., \cite{Cont2001RQUF}, numerous empirical studies in
opinion dynamics have not yet helped to establish even a basic set
of stylized facts, which could be seen as a goal for the theoretical
models.

In this contribution we will focus on the parties' vote share distribution
in a country as a proxy of the political attitude dynamics of the
country's population. Here we will focus on empirical data from the
Lithuanian parliamentary elections. Empirical data from the Lithuania
parliamentary elections was previously considered in many works by
Lithuanian political and social scientists, e.g., \cite{Degutis2000LitPolSci,Jurkynas2004JBaltStud,Ramonaite2006},
but in most of the approaches highly aggregated data was analyzed
and the analysis itself was just a means to an end. Numerous other
previous approaches have already considered empirical data gathered
during the various types of elections in the well-established democracies,
e.g., \cite{Fortunato2007PRL,Andersen2008IntJModPhysC,Borghesi2010EPJB,Mantovani2011EPL,Borghesi2012PLOS,FernandezGarcia2014PRL,Paz2015,Sano2016,Fenner2017QQ,Fenner2017}.
Different statistical features were considered in these and numerous
other papers, e.g., turnout distributions, spatial distributions,
open list ranking statistics, while some of the papers were dedicated
to the analysis of the vote share distributions. In
these different approaches different theoretical fits for the marginal
vote share distributions were proposed. In some of these works the
empirical vote share data was fitted using the log-normal distribution
\cite{Fortunato2007PRL}, the normal distribution \cite{Mantovani2011EPL,FernandezGarcia2014PRL},
distributions based on the Weibull distribution \cite{Paz2015,Fenner2017QQ}
and the beta distribution \cite{Sano2016,Fenner2017}. In this paper
we would like to argue that it may be hard to distinguish between
these distributions, but the beta distribution is likely to be the
best choice. A similar observation that numerous distribution may
fit the empirical statistics of religious populations was made in
\cite{Ausloos2007EPL}. It is worth to note that in order for these
statistical marginal distribution models to reasonably fit the data,
the respective multivariate distributions would be needed to be defined
on simplex as was done in some sophisticated approaches in political
science literature \cite{Aitchison1986,Katz1999APSR}. The Dirichlet
distribution, marginal distributions are the beta distributions, and
the logit-normal distribution, marginal distributions are similar
to the normal distribution, seem to acchieve this goal, which could
serve as an argument for their wider usage. But here in our analysis
we will limit ourselves to the comparison of the statistical models
of the marginal distributions. We will address the empirical analysis
from this point of view in Section~\ref{sec:Empirical-analysis}.

Most likely inspired by the theory of coarsening and the first-passage
phenomena numerous modelling approaches in sociophysics have considered
consensus formation in differently formulated models of the opinion
dynamics \cite{Clifford1973,Liggett1999,SznajdWeron2000IJMPC,SznajdWeron2005,Galam2008ModPhysC,Gunduc2015PhysA}.
These approaches are primarily interested in whether the agents will
reach the uniform consensus state or if the agent population will
remain heterogeneous in opinion. These models are usually based on
varying interpretations of the Ising model, such as the voter model
\cite{Clifford1973,Liggett1999}, Sznajd model \cite{SznajdWeron2000IJMPC,SznajdWeron2005}
or Galam model \cite{Galam2008ModPhysC}. Most of time these models
converge to the fixed states which are either consensus or coexistence
states. Yet it is well-known that opinion heterogeneity is rather
ubiquitous trait as well as that it is of rather dynamic nature. The
aforementioned models can be easily extended to include dynamism by
introducing exogeneous shocks, certain degree of contrarian behavior
or certain type of inflexibility into the model \cite{Galam2007PhysA,Deffuant2006JASSS,Nail2016APPA}.
Also most of these models are two-state models while in some cases
there are more than two viable options to choose from, e.g., usually
more than $15$ parties participate in Lithuanian parliamentary election
(with more than $4$ of them winning seats in the parliament by the
popular vote). In the political science and mathematical
literature one would find a more varied approaches \cite{Fara2014,Pukelsheim2014},
but usually their primary goal is to provide election procedures,
which would represent the opionion of the electorate the best. In
Section~\ref{sec:An-agent-based-model} of this paper we will discuss
an alternative possibility, based on Kirman's model \cite{Kirman1993QJE},
to formulate an agent-based model for the voting behavior.

This paper is divided into two main parts. In Section~\ref{sec:An-agent-based-model}
an agent-based model for the voting behavior is presented. In Section~\ref{sec:Empirical-analysis}
we discuss the empirical data and use the model from previous section
to reproduce the statistical patterns uncovered during the empirical
analysis. Finally we summarize our results and provide a discussion
in Section~\ref{sec:Discussion}.

\section{A multi-state agent-based model for the voting behavior\label{sec:An-agent-based-model}}

Originally in a seminal paper by Alan Kirman \cite{Kirman1993QJE}
a simple two-state herding behavior model was proposed. The aim of
the model was to reproduce similar behavioral patterns observed by
biologists and economists. It was noted that individuals tend to immitate
their peers' actions despite the lack of rational reasons to do so
\cite{Bass1969ManSci,Pastels1987Birkhauser1,Pastels1987Birkhauser2,Becker1991JPolitEco,Ishii2012NJP}.
This model is in some sense very much a like many other psychologically
motivated models \cite{Deffuant2006JASSS,Sobkowicz2012PlosOne,Lilleker2014,Sobkowicz2016PlosOne,Nail2016APPA,Duggins2017JASSS},
but, in comparison, this model is extremely simple and, as we will
show later, extremely efficient.

Kirman made an assumption that agents could change their behavior
on their own (acting according to the perceived attractiveness of
the available choices) or due to peer influence (recruitment mechanism).
In contemporary form this model is usually formulated using the one
step transition probabilities \cite{Alfarano2005CompEco,Alfarano2008Dyncon,Kononovicius2012PhysA}:
\begin{eqnarray}
P(X\rightarrow X+1) & = & \left(N-X\right)\left(\sigma_{1}+hX\right)\Delta t,\label{eq:two-state-1}\\
P\left(X\rightarrow X-1\right) & = & X\left[\sigma_{2}+h\left(N-X\right)\right]\Delta t,\label{eq:two-state-2}
\end{eqnarray}
here $N$ is a total number of agents in the modeled two-state system,
where each state represents different behavioral pattern (e.g., different
trading strategies in the financial market applications \cite{Alfarano2005CompEco,Alfarano2008Dyncon,Kononovicius2012PhysA}),
$X$ is a total number of agents occupying the first state (consequently
there are $N-X$ agents occupying the second state), $\sigma_{i}$
are the perceived attractiveness parameters, $h$ is an inter-agent
interaction intensity parameter, while $\Delta t$ is a relatively
short time step. In general it should be as small as possible, at
least so that a single agent could switch his state per time step.
In the scope of this paper $h$ parameter is not relevant, as here
we will not consider the temporal trends, so we can eliminate it by
introducing rescaled time $t_{s}=ht$ \cite{Alfarano2005CompEco,Alfarano2008Dyncon,Kononovicius2012PhysA}:
\begin{eqnarray}
P(X\rightarrow X+1) & = & \left(N-X\right)\left(\varepsilon_{1}+X\right)\Delta t_{s},\label{eq:two-state-1-1}\\
P\left(X\rightarrow X-1\right) & = & X\left[\varepsilon_{2}+\left(N-X\right)\right]\Delta t_{s},\label{eq:two-state-2-1}
\end{eqnarray}
here $\varepsilon_{i}=\frac{\sigma_{i}}{h}$ is the rescaled attractiveness
parameters. The dynamics of $x=\frac{X}{N}$, in the $N\rightarrow\infty$
limit, could be approximated by the Fokker\textendash Planck \cite{Aoki2007Cambridge,VanKampen2007NorthHolland}:
\begin{align}
\frac{\partial}{\partial t_{s}}p(x,t_{s}) & =-\frac{\partial}{\partial x}\left\{ \left[\varepsilon_{1}(1-x)-\varepsilon_{2}x\right]p(x,t_{s})\right\} +\frac{\partial^{2}}{\partial x^{2}}\left\{ x\left(1-x\right)p(x,t_{s})\right\} ,
\end{align}
or a stochastic differential equation \cite{Kononovicius2012PhysA}:
\begin{equation}
\rmd x\approx\left[\varepsilon_{1}(1-x)-\varepsilon_{2}x\right]\rmd t_{s}+\sqrt{2x(1-x)}\rmd W_{s}.
\end{equation}
From these equations it is rather straightforward to show that the
stationary distribution of $x$ is the beta distribution,$\mathcal{B}e\left(\varepsilon_{1},\varepsilon_{2}\right)$,
probability density function (PDF) of which is given by
\begin{equation}
p_{st}(x)=\frac{\Gamma(\varepsilon_{1}+\varepsilon_{2})}{\Gamma(\varepsilon_{1})\Gamma(\varepsilon_{2})}x^{\varepsilon_{1}-1}(1-x)^{\varepsilon_{2}-1}.
\end{equation}

The extension of the two-state model to describe switching between
multiple states is rather straightforward, though with some noteworthy
implications.

As the total number of agents, $\sum_{i}X_{i}=N$, is conserved, if
an agent switches his state, one state gains an agent, while the another
state looses an agent. With this in mind we can write the one step
transition probabilities, to and from state $i$, as follows:
\begin{equation}
P\left(X_{i}\rightarrow X_{i}\pm1\right)=\sum_{j\neq i}P\left(X_{i}\rightarrow X_{i}\pm1,X_{j}\rightarrow X_{j}\mp1\right),
\end{equation}
where the $P$ on the right hand side stands for the switching probability
between two different states. Let us assume that this $P$ takes the
same form as in the two-state model case, if so then we obtain:
\begin{eqnarray}
P(X_{i}\rightarrow X_{i}+1) & = & \sum_{j\neq i}X_{j}\left(\sigma_{ji}+h_{ji}X_{i}\right)\Delta t,\label{eq:multi-state-1}\\
P\left(X_{i}\rightarrow X_{i}-1\right) & = & X_{i}\sum_{j\neq i}\left[\sigma_{ij}+h_{ij}X_{j}\right]\Delta t.\label{eq:multi-state-2}
\end{eqnarray}

Although the current form of the transition probabilities allows some
flexibility, but the analytical treatment of the multi-state model
seems to be impossible as the one step transition probabilities for
$X_{i}$ depend on other $X_{j}$ ($j\neq i$) in non-trivial manner.
To eliminate this cumbersome dependence let us assume that:
\begin{itemize}
\item the perceived attractiveness of a state, $\sigma_{ij}$, does not
depend on agent from which state is attracted to it, $\sigma_{ij}=\sigma_{j}$;
\item the interaction intensity is symmetric and independent of the states
interacting agents are in, $h_{ij}=h$.
\end{itemize}
Note that these assumptions are the opposite of what is assumed by
the well-known bounded confidence model \cite{Deffuant2006JASSS}.
Yet these assumptions allows us to further simplify the one step transition
probabilities:
\begin{eqnarray}
P\left(X_{i}\rightarrow X_{i}+1\right) & = & \left(N-X_{i}\right)\left(\varepsilon_{i}+X_{i}\right)\Delta t_{s},\label{eq:np-state-1}\\
P\left(X_{i}\rightarrow X_{i}-1\right) & = & X_{i}\left(\varepsilon_{-i}+N-X_{i}\right)\Delta t_{s},\label{eq:np-state-2}
\end{eqnarray}
here $\varepsilon_{-i}=\sum_{j\neq i}\varepsilon_{j}$ is the total
attractiveness of switching away from $i$. Because these switching
probabilities have the same form as Eqs. (\ref{eq:two-state-1-1})
and (\ref{eq:multi-state-1}), the stationary distribution of $x_{i}$
is most likely to be $\mathcal{B}e\left(\varepsilon_{i},\varepsilon_{-i}\right).$
In that case the stationary distribution of $\boldsymbol{x}$ is $\Dir\left(\boldsymbol{\varepsilon}\right)$.
Though note that if the simplifying assumptions are violated the stationary
distribution of $\boldsymbol{x}$ might no longer be the Dirichlet
distribution, nor the marginal distribution of at least some of the
$x_{i}$ might no longer follow the beta distribution.

In the previous paragraphs we have defined the two-state and multi-state
models ignoring the underlying interaction topology, or namely we
defined the models as a mean-field models. Yet these models can be
easily generalized to take interaction topologies, e.g, some kind
of random networks \cite{Alfarano2009Dyncon,Kononovicius2014EPJB},
into account. In such case one has to define individual agent switching
probabilities:
\begin{equation}
P_{a}\left(O\rightarrow D\right)=\left[\varepsilon_{D}+n_{a}(D)\right]\Delta t_{s},\label{eq:prob-agent-transition}
\end{equation}
here $a$ is an index, which identifies individual agents, $O$ and
$D$ represent origin and destination states respectively ($O\neq D$),
while $n_{a}(D)$ is a number of agent's $a$ neighbors who are in
state $D$. As long as average degree of nodes on the network is large,
i.e., comparable with the total number of agents, links are uncorrelated
and $\Delta t_{s}$ is small, both the discussed mean-field models
and this model should produce the same results \cite{Kononovicius2014EPJB}.
If the average degree is significantly smaller than the total number
of agents, one can still use one-step transition probabilities to
describe the system at macro-level, but the probabilities will be
of a slightly different form than Eqs.~(\ref{eq:np-state-1}) and
(\ref{eq:np-state-2}). Here we would like to note that if we set
$\varepsilon_{D}=0$ and restrict the model to two states, then Eq.\ (\ref{eq:prob-agent-transition})
basically describes the dynamics underlying the well known Voter model
\cite{Clifford1973,Liggett1999,Castellano2009RevModPhys}.

\section{Empirical analysis of the Lithuanian parliamentary elections\label{sec:Empirical-analysis}}

Let us start with a brief description of the two-tier voting system
used during the Lithuanian parliamentary elections. The elections
are held quadrennially (the exact date is set by the president of
Lithuania). During every parliamentary election all of the seats in
the parliament are being contested. Namely, some members of the parliament
may serve a shorter than a four year term, if they have replaced somebody
else. $71$ of the seats in the parliament are allocated to the elected
representatives of the $71$ electoral districts (two-round system
is used to elect the representatives of the electoral districts).
The remaining $70$ seats are distributed according to the popular
vote for an open party list. The party needs to pass the threshold
of $5\%$ of the popular vote to obtain at least $1$ seat in this
way.

From a voters perspective the parliamentary elections takes form of
two ballots. One ballot is used to vote for a representative of the
electoral district. Usually the district representative is not elected
in the first round and the second round is held. This time the voter
will be able to from a list of the top two candidates from the first
round. Another ballot is used to vote for a political party or movement
and optionally for the up to $5$ people from the respective party's
or movement's list.

In this paper we are concerned with the distribution of popular vote
for the parties (from here onwards we will refer to both parties and
movements simply as parties) across all of the polling stations. Each
of the $71$ electoral districts has multiple polling stations (there
are usually $\sim2000$ polling stations in total). Every voter is
assigned to one of the polling stations based on the location of residence.
Due to uneven population density and consideration of the polling
station proximity to the voters, the polling stations vary in the
number of the assigned voters \textendash{} some of the smallest polling
stations could have as few as $100$ assigned voters, while the largest
could have $7000$. In this contribution we will ignore this difference
and consider each polling station as a single unit providing one data
point per party per election. In our analysis we ignore voting by
mail as well as voting in polling stations abroad in order to at least
to some extent ensure that the polling stations are spatially separate,
namely to keep chances that person lives and interacts with people
from the other polling station as low as it is possible.

The data used in this analysis is freely available at \url{https://www.rinkejopuslapis.lt/ataskaitu-formavimas}
(the website is managed by the Central Electoral Commission of the
Republic of Lithuania). Sadly at the time of writing the website works
well in Lithuanian language only (some parts remain untranslated).
Therefore we have provided a cleaned up version of the data used in
our analysis at \url{https://github.com/akononovicius/lithuanian-parliamentary-election-data}
(the original data sets were downloaded on August 31, 2016).

Let us start the empirical analysis by considering the Lithuanian
parliamentary election of 1992. $17$ parties competed in that election,
but only $4$ of them won seats by the popular vote. One of the four
parties just barely made past the threshold of $5\%$, while the other
three enjoyed noticeably larger support. As the fourth party introduces
distorting effect, which we will discuss a bit later, let us ignore
it as well as the other $13$ parties, which did not make past the
threshold. The statistical properties of the remaining three most
popular parties are not significantly distorted by various additional
effects, so let us consider their renormalized vote share for the
introductory analysis. We use the following party name abbreviations
for these parties: SK will stand for ``S\k{a}j\={u}džio koalicija'',
LKDP \textendash{} ``Lietuvos krikš\v{c}ioni\k{u} demokrat\k{u} partijos,
Lietuvos politini\k{u} kalini\k{u} ir tremtini\k{u} s\k{a}jungos ir
Lietuvos demokrat\k{u} partijos jungtinis s\k{a}rašas'', LDDP \textendash{}
``Lietuvos demokratin\.{e} darbo partija''.

\begin{figure}
\noindent \begin{centering}
\includegraphics[width=0.7\textwidth]{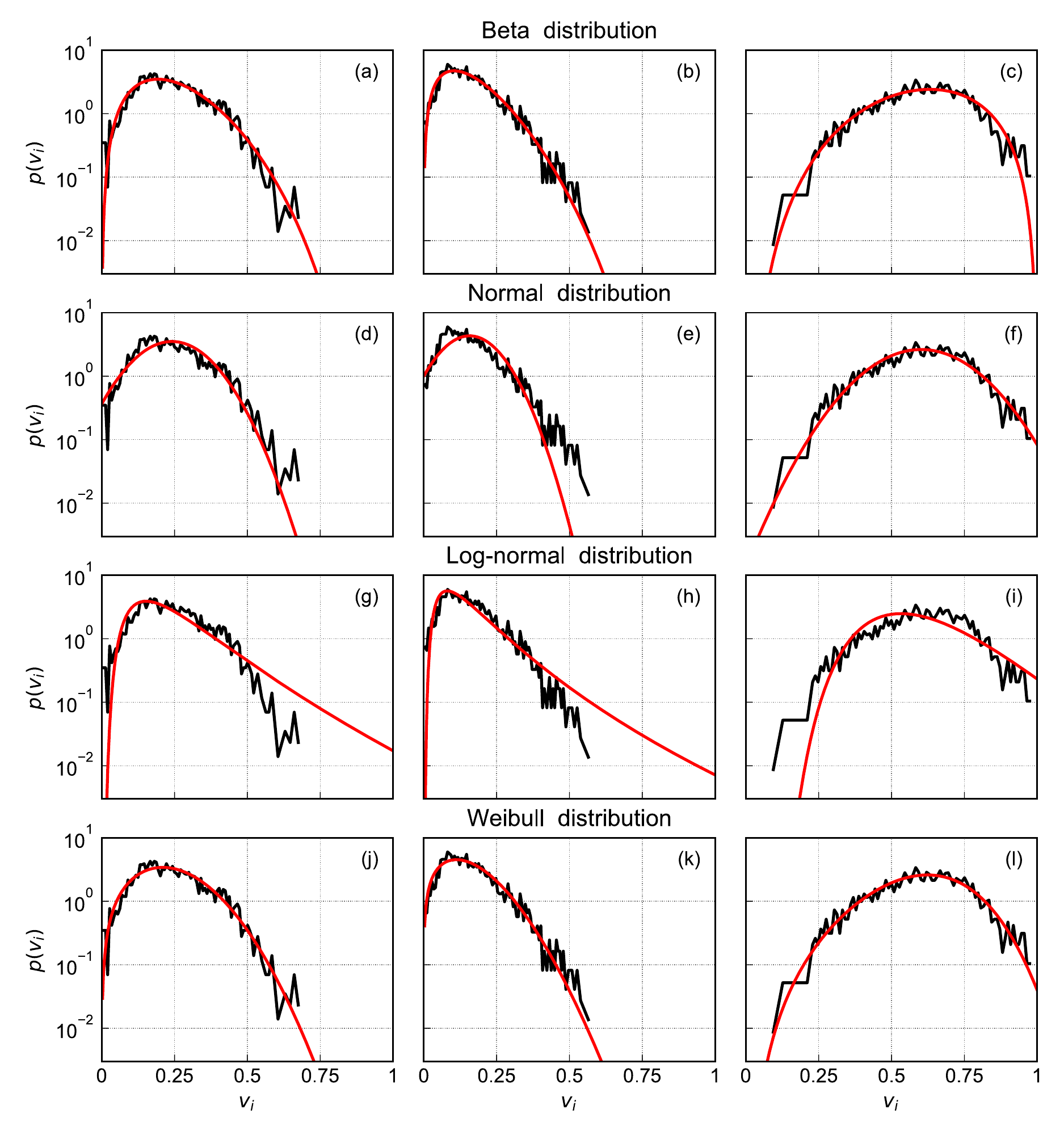}
\par\end{centering}
\caption{(color online) The empirical vote share PDFs (black curves) fitted
by the four commonly used distributions (red curves): (a)-(c) the
beta distribution, (d)-(f) the normal distribution, (g)-(i) the log-normal
distribution and (j)-(l) the Weibull distribution. Renormalized empirical
vote shares of the three most popular parties during the 1992 election
were used: SK ((a), (d), (g), (j)), LKDP ((b), (e), (h), (k)) and
LDDP ((c), (f), (i), (l)). The respective parameter values are given
in Table~\ref{table:empir-dist-params-1992}.}

\label{fig:empir-dist-fits-1992}
\end{figure}

\begin{figure}
\noindent \begin{centering}
\includegraphics[width=0.4\textwidth]{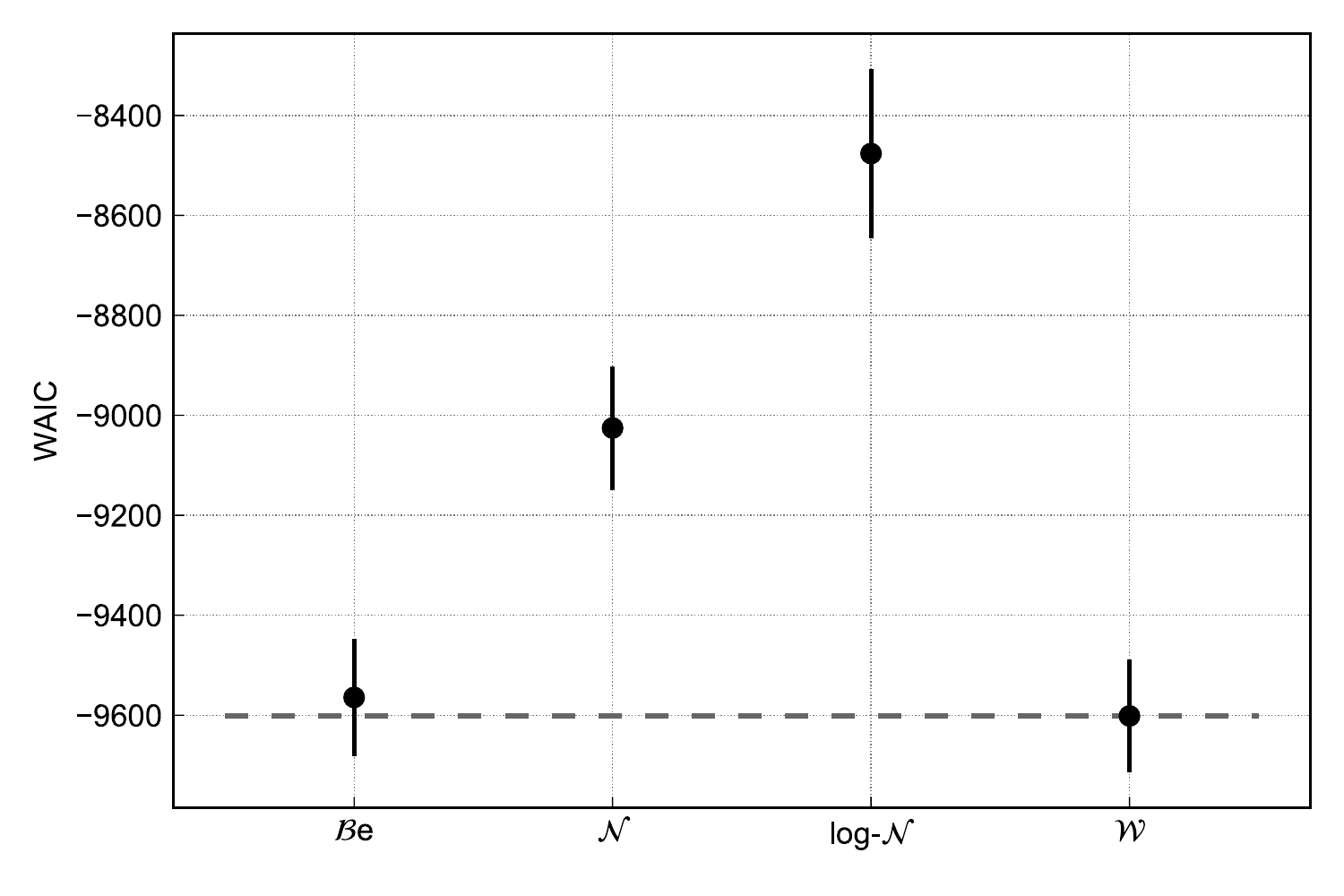}
\par\end{centering}
\caption{Comparison of the suitability of the four commonly used distributions
as reflected by WAIC in case of the empirical data from the 1992 parliamentary
election. The dashed line represents best (lowest) WAIC value (the
Weibull distribution WAIC).}

\label{fig:empir-dist-fit-comparison-1992}
\end{figure}

In Fig.~\ref{fig:empir-dist-fits-1992} we show the vote share PDFs
of SK, LKDP and LDDP fitted by the four distributions commonly used
in the literature \cite{Ausloos2007EPL,Fortunato2007PRL,Mantovani2011EPL,FernandezGarcia2014PRL,Paz2015,Sano2016,Fenner2017QQ,Fenner2017}.
The respective distribution parameters are given in Table~\ref{table:empir-dist-params-1992}.
As can be easily seen from the figure the beta and Weibull distributions
provide good fits for the empirical PDFs, normal distribution also
provides a rather good fit, while log-normal distribution seems to
be somewhat off. A formal comparison of the relative quality of these
fits can be checked using Watanabe\textendash Akaike information criterion
(abbr. WAIC) \cite{Watanabe2013}. These values are given alongside
parameter values in Table~\ref{table:empir-dist-params-1992}. See
Fig.~\ref{fig:empir-dist-fit-comparison-1992}, for a visual comparison
between the obtained WAIC values. From the table and the figure it
can be seen that the fits provided by the beta and Weibull distributions
are of similar quality, with the fit provided by the Weibull distribution
being slightly better. While the fits provided by the normal and log-normal
distributions have noticeably larger WAIC values and thus provide
relatively worse fits for the data.

\begin{table}
\caption{The selected distribution parameter values, inferred while analyzing
the vote shares of the three most popular parties during the 1992
parliamentary election, and the respective WAIC values.}

\noindent \begin{centering}
\begin{tabular}{|c|c|c|c|c|}
\hline 
\multirow{3}{*}{$\mathcal{B}e$} & SK & $\alpha=3.08\pm0.18$ & $\beta=9.73\pm0.55$ & \multirow{3}{*}{$WAIC=-9560\pm120$}\tabularnewline
\cline{2-4} 
 & LKDP & $\alpha=2.32\pm0.14$ & $\beta=12.44\pm0.75$ & \tabularnewline
\cline{2-4} 
 & LDDP & $\alpha=5.45\pm0.35$ & $\beta=3.6\pm0.2$ & \tabularnewline
\hline 
\multirow{3}{*}{$\mathcal{N}$} & SK & $\mu=0.241\pm0.005$ & $\sigma=0.113\pm0.003$ & \multirow{3}{*}{$WAIC=-9020\pm120$}\tabularnewline
\cline{2-4} 
 & LKDP & $\mu=0.157\pm0.004$ & $\sigma=0.092\pm0.002$ & \tabularnewline
\cline{2-4} 
 & LDDP & $\mu=0.602\pm0.005$ & $\sigma=0.152\pm0.004$ & \tabularnewline
\hline 
\multirow{3}{*}{log-$\mathcal{N}$} & SK & $\mu=-1.556\pm0.028$ & $\sigma=0.57\pm0.02$ & \multirow{3}{*}{$WAIC=-8480\pm170$}\tabularnewline
\cline{2-4} 
 & LKDP & $\mu=-2.05\pm0.03$ & $\sigma=0.69\pm0.02$ & \tabularnewline
\cline{2-4} 
 & LDDP & $\mu=-0.547\pm0.013$ & $\sigma=0.289\pm0.008$ & \tabularnewline
\hline 
\multirow{3}{*}{$\mathcal{W}$} & SK & $k=2.25\pm0.07$ & $\lambda=0.272\pm0.006$ & \multirow{3}{*}{$WAIC=-9600\pm110$}\tabularnewline
\cline{2-4} 
 & LKDP & $k=1.79\pm0.05$ & $\lambda=0.177\pm0.005$ & \tabularnewline
\cline{2-4} 
 & LDDP & $k=4.53\pm0.15$ & $\lambda=0.658\pm0.006$ & \tabularnewline
\hline 
\end{tabular}
\par\end{centering}
\label{table:empir-dist-params-1992}
\end{table}

\begin{figure}
\noindent \begin{centering}
\includegraphics[width=0.4\textwidth]{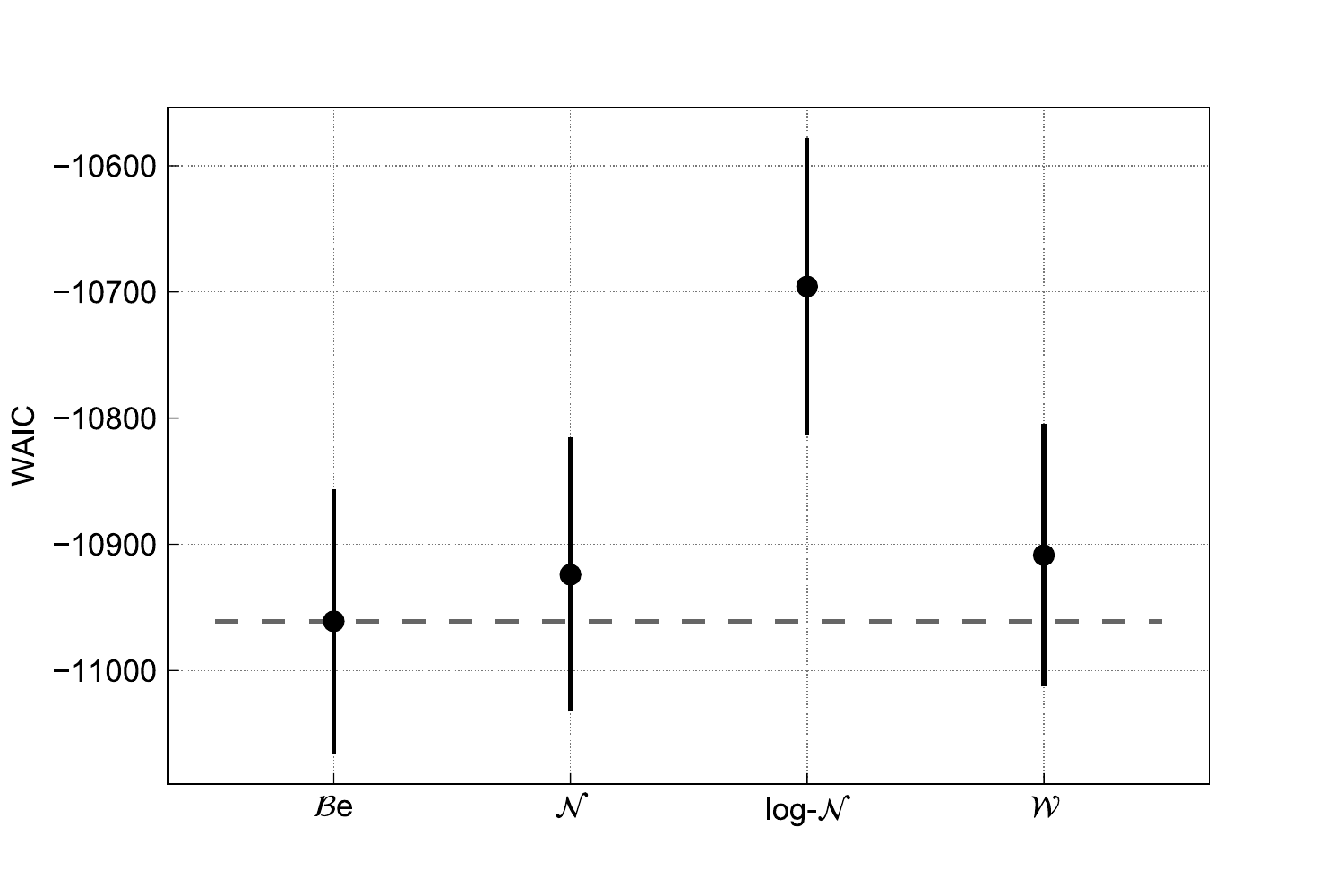}
\par\end{centering}
\caption{Comparison of the suitability of the four commonly used distributions
as reflected by WAIC in case of the surrogate data (true distribution
is the multivariate beta distribution). The dashed line represents
best (lowest) WAIC value of the four (the beta distribution WAIC).}

\label{fig:surrogate-data}
\end{figure}

In the literature the log-normal distribution was used to model the
vote shares of individual politicians on the open party lists \cite{Fortunato2007PRL}.
We consider different data so it is not strange at all that the log-normal
distribution provides the worst fit. Another difference from the approach
taken in \cite{Fortunato2007PRL} is that we use different normalization
procedure. Namely we renormalize a sum of the vote shares of the considered
parties to be equal to $1$ in all polling stations, instead of dividing
the vote share by the mean for the respective party. While the normal
distribution was quite successfully used for the data gathered during
various elections held in well-established democracies \cite{Mantovani2011EPL,FernandezGarcia2014PRL}.
We believe that normal distribution seems to provide a good fit in
those cases due to similar vote shares received by the competing parties
(the mean vote shares are of similar magnitude) as well as smaller
variability of the vote share between the polling stations (smaller
standard deviation). This can be also captured by the beta distribution
assuming that $\alpha$ and $\beta$ are large, which
would reconcile these results with our approach and the approaches
found in \cite{Sano2016,Fenner2017}. So let us confirm this intuition
by generating surrogate data, distributed according to the Dirichlet
distribution (the multivariate beta distribution) with $\boldsymbol{\alpha}=\left\{ 10,10,10\right\} $.
As can be seem from Fig.~\ref{fig:surrogate-data}, the normal distribution
provides a relatively good fit for the generated surrogate data, while
the beta distribution and the Weibull distribution also provide a
relatively good fits. In the context of the model defined in the previous
section, this would mean that in countries with older democratic traditions
self-induced transitions are significantly more common than peer-induced
transitions.

\begin{figure}
\noindent \begin{centering}
\includegraphics[width=0.9\textwidth]{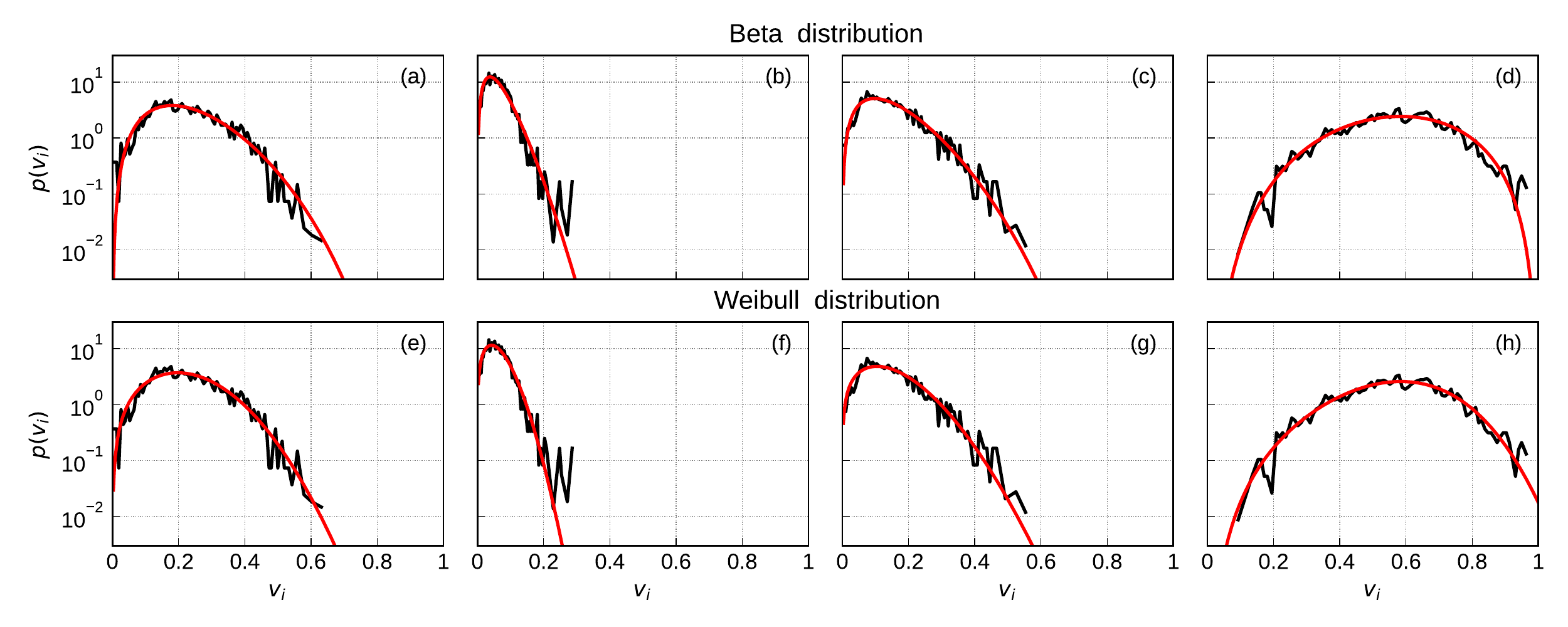}
\par\end{centering}
\caption{(color online) The empirical vote share PDFs (black curves) fitted
by the (a)-(d) beta and (e)-(h) Weibull distributions (red curves).
Renormalized empirical vote shares of the four most popular parties
during the 1992 election were used: SK ((a) and (e)), LSDP ((b) and
(f)), LKDP ((c) and (g)) and LDDP ((d) and (h)). The respective parameter
values are given in Table~\ref{table:empir-dist-params-1992-extended}.}

\label{fig:empir-dist-fits-1992-1}
\end{figure}

Now lets extend the initial analysis by including the fourth party
which passed the $5\%$ threshold, ``Lietuvos socialdemokrat\k{u}
partija'' (abbr. LSDP). At this point we would also like to stop
using the normal and log-normal distributions as they do not seem
to provide good fits for the reasons discussed previously. As you
can see from Fig.~\ref{fig:empir-dist-fits-1992-1} and Table~\ref{table:empir-dist-params-1992-extended}
both the beta and Weibull distributions provide similarly good fits
for the data. Note that parameter values $\beta$ (for the beta distribution)
and $\lambda$ (for the Weilbull distribution) inferred for the LSDP
are of somewhat different magnitude than the rest of parameter values.
In the context of the model defined in the previous section we could
see this as voters choosing to vote for another ``like-able'' party,
because of lack of belief that the preferred party could win enough
votes in the election. LDDP was another left-wing party in the 1992
election, which the most likely attracted a significant share of LSDP
voters.

\begin{table}
\caption{The selected distribution parameter values, inferred while analyzing
the vote shares of the four most popular parties during the 1992 parliamentary
election, and the respective WAIC values.}

\noindent \begin{centering}
\begin{tabular}{|c|c|c|c|c|}
\hline 
\multirow{4}{*}{$\mathcal{B}e$} & SK & $\alpha=3.25\pm0.19$ & $\beta=9.73\pm0.55$ & \multirow{4}{*}{$WAIC=-18400\pm170$}\tabularnewline
\cline{2-4} 
 & LSDP & $\alpha=2.42\pm0.14$ & $\beta=37.1\pm2.4$ & \tabularnewline
\cline{2-4} 
 & LKDP & $\alpha=2.37\pm0.13$ & $\beta=13.7\pm0.7$ & \tabularnewline
\cline{2-4} 
 & LDDP & $\alpha=5.42\pm0.31$ & $\beta=4.15\pm0.24$ & \tabularnewline
\hline 
\multirow{4}{*}{$\mathcal{W}$} & SK & $k=2.28\pm0.09$ & $\lambda=0.254\pm0.005$ & \multirow{4}{*}{$WAIC=-18450\pm170$}\tabularnewline
\cline{2-4} 
 & LSDP & $k=1.76\pm0.06$ & $\lambda=0.069\pm0.002$ & \tabularnewline
\cline{2-4} 
 & LKDP & $k=1.78\pm0.06$ & $\lambda=0.166\pm0.005$ & \tabularnewline
\cline{2-4} 
 & LDDP & $k=4.24\pm0.14$ & $\lambda=0.622\pm0.007$ & \tabularnewline
\hline 
\end{tabular}
\par\end{centering}
\label{table:empir-dist-params-1992-extended}
\end{table}

\begin{figure}
\noindent \begin{centering}
\includegraphics[width=0.7\textwidth]{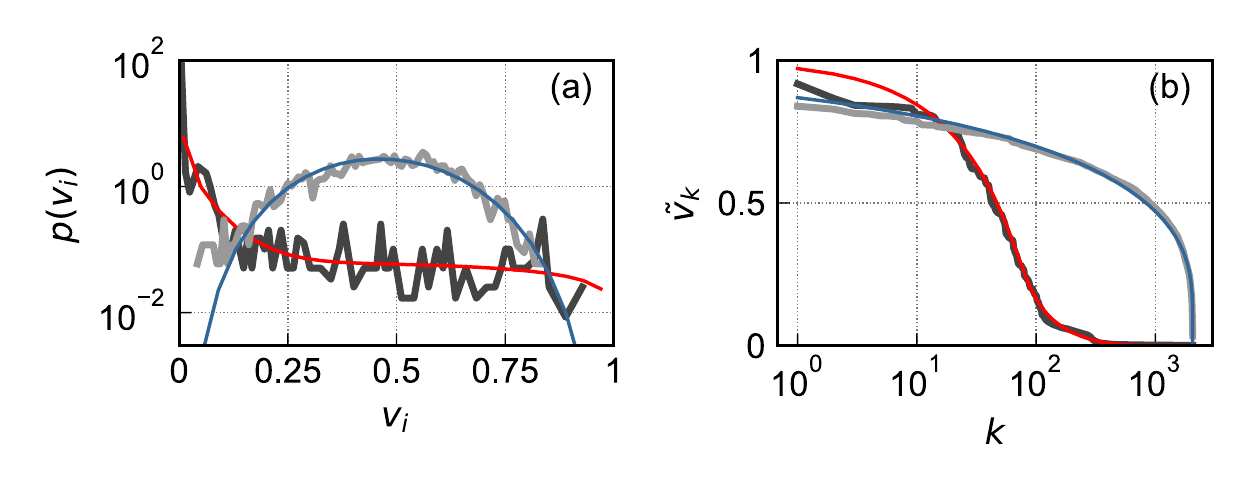}
\par\end{centering}
\caption{(color online) Comparison between vote share PDFs (a) and rank-size
distribution (b) of LLS (pronounced vote segregation effect) and LDDP
(without pronounced vote segregation effect) during the 1992 parliamentary
election. Empirical data shown as wide gray curves (dark \textendash{}
LLS, light \textendash{} LDDP), while best fits shown as narrow colored
curves. Best fits are provided by a mixture of beta distributions
(red curve), $0.95\cdot\mathcal{B}e\left(0.08,10\right)+0.05\cdot\mathcal{B}e\left(1.22,1.37\right)$,
and the beta distribution (blue curve), $\mathcal{B}e\left(5.7,6.5\right)$.}

\label{fig:lls-segregation}
\end{figure}

Let us now consider another less popular party named ``Lietuvos lenk\k{u}
s\k{a}junga'' (abbr. LLS). This party is interesting to us as its
vote share distribution exhibits another interesting effect \textendash{}
vote segregation effect. We would like to claim that this is related
to the fact that LLS was mainly supported by the ethnic minorities,
which are geographically segregated (most of the ethnic minorities
living in major cities and Vilnius County). This creates a need to
use a mixture distribution, because distribution parameter values
will be different for those regions in which ethnic minorities make
up a significant part of population and for those regions in which
representatives of ethnic minorities are few. In Fig.~\ref{fig:lls-segregation}
we have compared the vote share and rank-size distributions of the
two parties from the 1992 parliamentary election: one with pronounced
segregation effect (LLS) and one without pronounced segregation effect
(LSDP). LSDP vote share distribution seems to be rather well fitted
by the beta distribution, while LLS vote share distribution is fitted
using a mixture of the two beta distributions.

\begin{figure}
\noindent \begin{centering}
\includegraphics[width=0.7\textwidth]{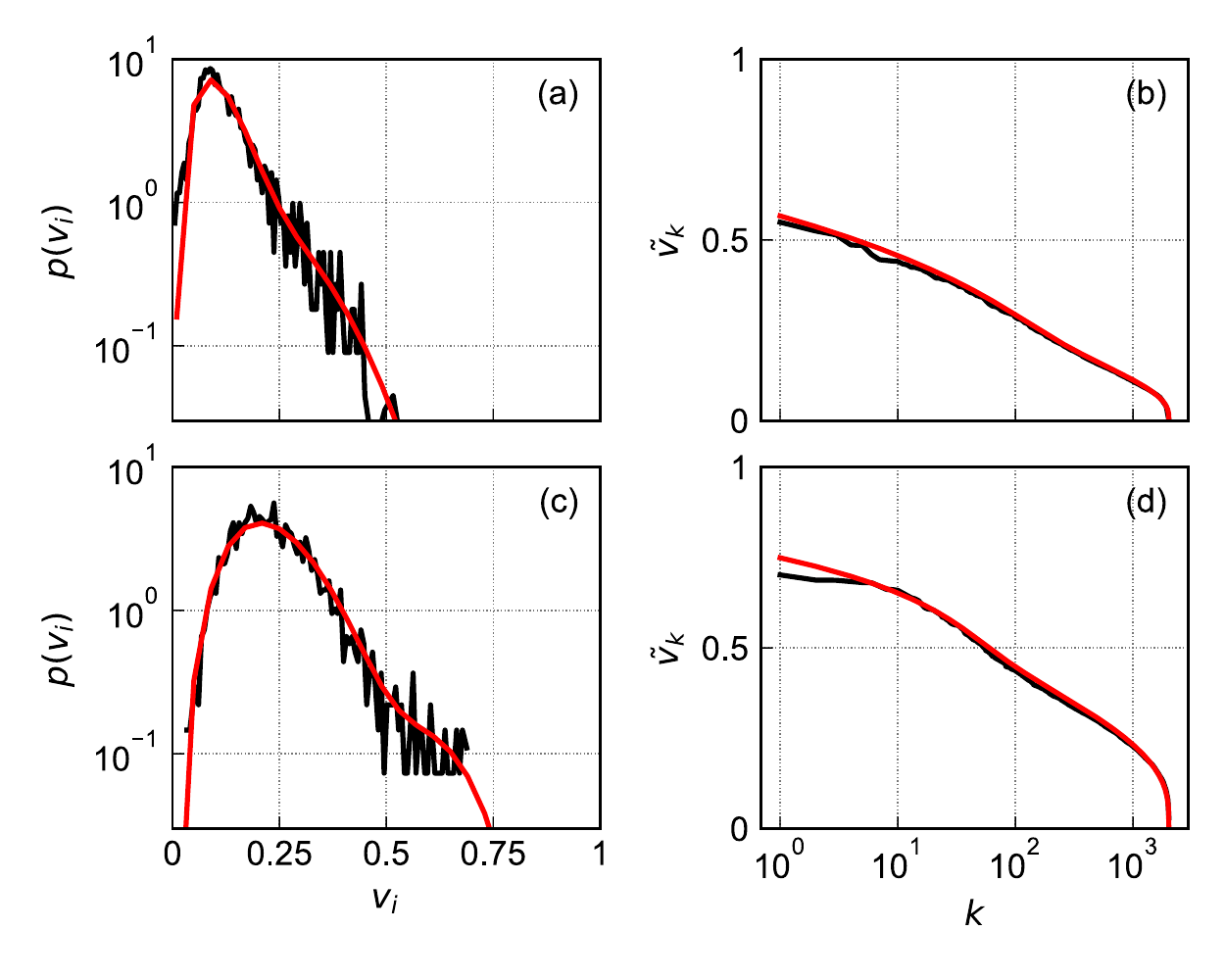}
\par\end{centering}
\caption{(color online) The vote share PDFs ((a) and (c)) and the rank-size
distribution ((b) and (d)) of LSDP during the 2008 election ((a) and
(b)) and ``Darbo partija'' during the 2012 election ((c) and (d)).
Empirical statistical properties are shown as black curves, while
fits using a mixture of the beta distributions are shown in red. Parameter
values of the fitting distributions were set as follows: $0.85\cdot\mathcal{B}e\left(3.9,31.7\right)+0.15\cdot\mathcal{B}e(4.3,12.9)$
(for LSDP), $0.97\cdot\mathcal{B}e\left(4.5,14.5\right)+0.03\cdot\mathcal{B}e\left(15.3,11.1\right)$
(for ``Darbo partija'').}

\label{fig:later-segregation}
\end{figure}

While analyzing the data from the later elections, we find that the
other parties also start to exhibit vote segregation patterns. For
a more recent examples of this pattern see Fig.~\ref{fig:later-segregation}.
This could be a sign that a party is finding its political niche,
taking over a segment of electorate (e.g., by promoting policies favoring
certain socio-demographic groups). Also this could be a sign that
a party is taking over municipalities (e.g., by pressuring public
sector workers to vote for certain party with a threat to lose their
jobs).

Let us now use the insights from the previous paragraphs to select
parameters for the model proposed in previous section. Our aim is
to reproduce the 1992 parliamentary election vote share distributions.
We consider SK, LKDP and LDDP as well as the ``Other'' party (abbr.
O), which is composed of the other $14$ parties which participated
in that election. The ``Other'' party includes LSDP, so we will
have to violate the simplifying assumptions, and LLS, so we will have
to pick two different parameter sets: one for the polling stations
in which LLS was weak ($\sim95\%$ of polling stations) and the other
for the polling stations in which LLS was strong ($\sim5\%$ of polling
stations). Initial parameter values were estimated using Bayesian
inference and later adjusted to obtain a better fit between the model
and the empirical data, after the fitting procedure we arrived at
the following parameter set:
\begin{equation}
\varepsilon^{(95\%)}=\left(\begin{array}{cccc}
0 & 2.6 & 9.3 & 4.7\\
3.8 & 0 & 9.3 & 4.7\\
3.8 & 2.6 & 0 & 4.7\\
3.8 & 2.6 & 15.3 & 0
\end{array}\right),\qquad\varepsilon^{(5\%)}=\left(\begin{array}{cccc}
0 & 0.35 & 2.5 & 7\\
0.65 & 0 & 2.5 & 7\\
0.65 & 0.35 & 0 & 7\\
0.65 & 0.35 & 2.5 & 0
\end{array}\right).\label{eq:model-params}
\end{equation}

\begin{figure}
\noindent \begin{centering}
\includegraphics[width=0.7\textwidth]{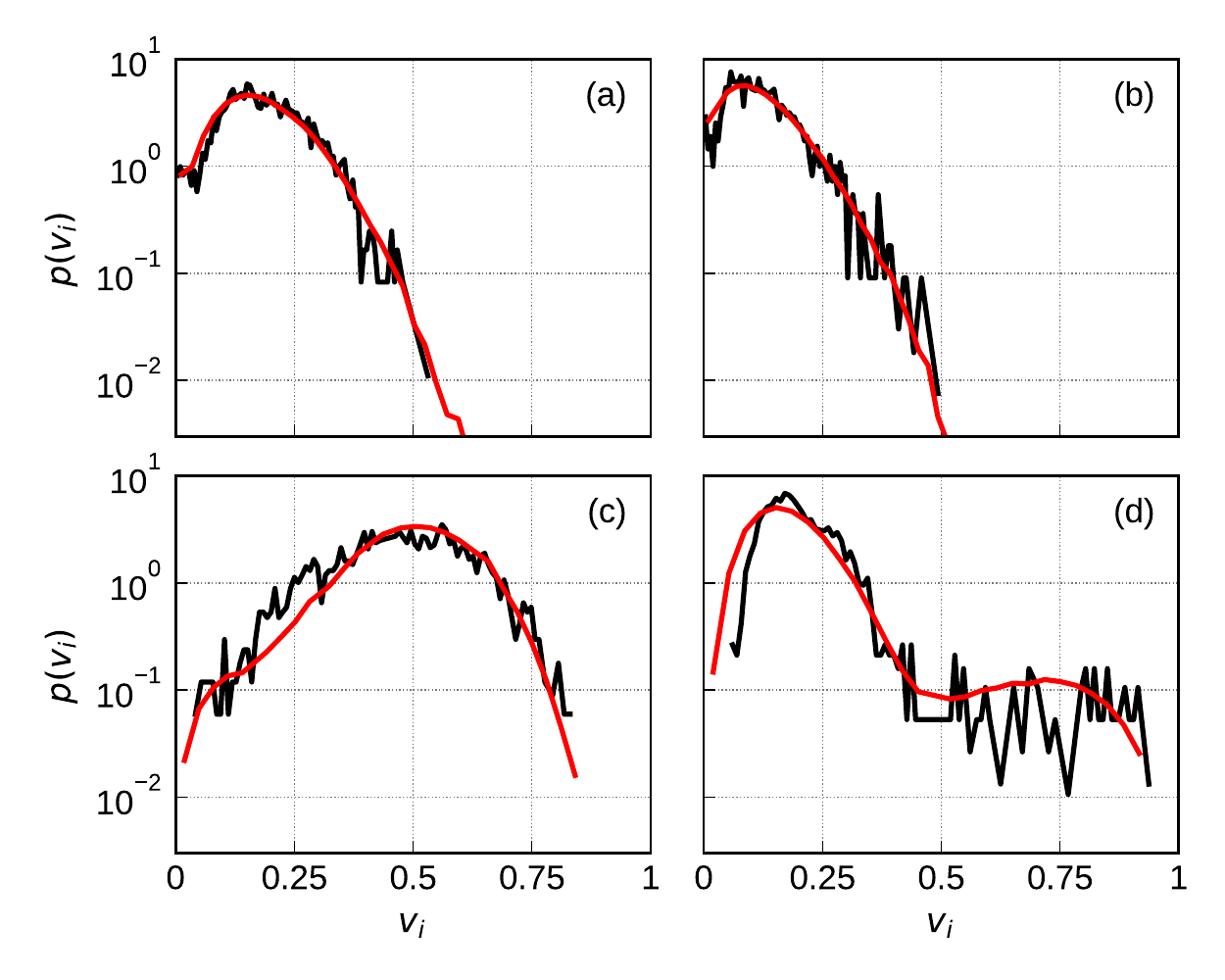}
\par\end{centering}
\caption{(color online) Reproducing the vote share PDFs of (a) SK, (b) LKDP,
(c) LDDP and (d) O observed during the 1992 parliamentary election.
Black curves show the empirical PDFs, while red curves were obtained
by numerically simulating the agent-based model discussed in Section~\ref{sec:An-agent-based-model}.
The used model parameter values are given in Eq.~(\ref{eq:model-params}).}

\end{figure}

\section{Conclusions\label{sec:Discussion}}

In this paper we have considered the vote share distributions observed
in the Lithuanian parliamentary election data. Most of the attention
was given to thorough analysis of the 1992 parliamentary election
data set. We have compared four competing distributions often used
to fit various election data sets from around the world \cite{Ausloos2007EPL,Fortunato2007PRL,Mantovani2011EPL,FernandezGarcia2014PRL,Paz2015,Sano2016,Fenner2017QQ,Fenner2017}
and found that the beta and Weibull distributions seem to provide
good fit, while normal and log-normal distributions are ill-suited
if peer influence is strong in comparison to the independent voter
behavior. We have also shown that as democratic tradition takes root
parties start to take over electoral segments. As these segments are
usually differentiated based on socio-economic properties, the voters
belonging to different segments are segregated and thus vote share
distributions also become segregated. Segregated vote share distributions
are well fitted by a mixture of the beta distributions (although they
could be also well fitted by a mixture of Weibull distributions \cite{Fenner2017QQ}).

To provide sound argument for the use of the beta distributions we
have formulated a multi-state agent-based model for the voting behavior.
This model, under certain simplifying assumptions, produces Dirichlet
distributed vote share (marginal distributions of which are distributed
according to the beta distribution). One could violate these assumptions
to add some flexibility to the model. Note that proposed model is
very different from the psychologically motivated models for the opinion
dynamics or voting behavior, e.g., bounded confidence model \cite{Deffuant2006JASSS},
thus we would like to raise the idea that vote share distribution
could reflect some other processes in addition to (or instead of)
opinion dynamics. A similar idea was already raised in \cite{FernandezGarcia2014PRL}.
This suggests a possible future development for both the agent-based
modeling and empirical analysis \textendash{} to consider spatio-temporal
modeling of the Lithuania parliamentary elections.

\end{document}